%% file: main.tex
\documentclass[twocolumn, table]{aastex63}
\usepackage{xcolor}
\usepackage[utf8]{inputenc}
\usepackage{comment}
\usepackage{gensymb}

\input{header}

\received{June 1, 2019}
\revised{January 10, 2019}
\accepted{\today}
\submitjournal{RNAAS}

\shorttitle{Jet Structure}
\shortauthors{Tandon \& Lloyd-Ronning}
\graphicspath{{./}{figures/}}

\begin{document}

\title{Understanding Gamma-ray Burst Jet Structures from Afterglow Light Curves}

\author[0000-0002-7624-4883]{Celia R. Tandon}
\affiliation{Stanford University, Palo Alto, CA 94305}
\affiliation{CCS-2, Computational Physics and Methods, Los Alamos National Lab, Los Alamos NM 87544}
\author[0000-0003-1707-7998]{Nicole M. Lloyd-Ronning}
\affiliation{CCS-2, Computational Physics and Methods, Los Alamos National Lab, Los Alamos NM 87544}
\affiliation{Department of Math, Science and Engineering, University of New Mexico, Los Alamos, Los Alamos NM 87544}




\begin{abstract}

Gamma-ray bursts (GRBs), associated with the collapse of massive stars or the collisions of compact objects, are the most luminous events in our universe. However, there is still much to learn about the nature of the relativistic jets launched from the central engines of these objects.  We examine how jet structure—that is, the energy and velocity distribution as a function of angle—affects observed GRB afterglow light curves. Using the package afterglowpy, we compute light curves arising from an array of possible jet structures, and present the suite of models that can fit the coincident electromagnetic observations of GW190814 (which is likely due to a background AGN). Our work emphasizes not only the need for broadband spectral and timing data to distinguish among jet structure models, but also the necessity for high resolution radio follow-up to help resolve background sources that may mimic a GRB afterglow.  
\\ \\

\end{abstract}



\section{Introduction}
\label{sec:intro}


Gamma-Ray Bursts (GRBs) are brief (less than a second to hundreds of seconds) but extremely luminous flashes of gamma-rays created by internal dissipation processes in ultra relativistic jets launched from central compact object-disk systems. As the jet front ploughs into the external medium and sweeps up matter, it decelerates and produces long-lived afterglow emission across the entire electromagnetic spectrum \cite{Pir04}. Many important details of relativistic jets—how they are launched, the internal processes that lead to the gamma-ray emission, the degree of magnetization, the physics of their collimation—are yet to be understood.

For short GRBs in particular, given the increased detection of coincident gravitational wave (GW) emission from their progenitor systems (the merger of two compact objects), the broadband GRB afterglow combined with gravitational wave emission can provide deep insight into the physics of their progenitors.


One of the major open questions in understanding the emission from electromagnetic counterparts to gravitational wave events is getting a handle on the so-called structure of the jet.  It has long been recognized that a GRB jet will have some sort of angular structure—that is a distribution of the jet outflow velocity and energy as a function of angle from the jet axis\footnote{The jet will also have radial structure, but here we consider emission from the jet front at roughly a singular radius}. \cite{Urr21} and \cite{Mur21}, for example, both show that the jet interaction with the wind can produce complicated jet structures, including an inverted like structure.

 With the propsect of continued, improved LIGO observations as well as a number of additional GW detectors slated to come online in the next few years, the need for a solid understanding of potential electromagnetic counterparts to GW events is stronger than ever. 
Below, we show how the angular structures of a jet affect the observed afterglow light curves.  We use the electromagnetic emission coincident with GW190814 (likely a background AGN \citep{alexander}) to illustrate how a number of physically plausible GRB jet models can reproduce the data. Our work emphasizes the need not only for broadband spectral and timing modelling, but the necessity of high resolution radio follow-up to rule out contamination from background sources.

\section{Jet Models} \label{jet}

 
\subsection{Jet Structures} 
    A structured jet is a GRB jet model where the isotropic-equivalent energy, $E_{iso}$, of the blast wave is a function of the angle from the jet axis:

\begin{equation}
    E_{iso} = 4 \pi \frac{dE}{d\Omega}
\end{equation}
The particular angular structure of the jet is first determined by the jet launching mechanism and then modified as the jet burrows out of the surrounding ejecta debris or stellar envelope \citep{afterglowpy}.

Below are some common jet structures found in the literature. These models are parameterized by a normalization $E_{iso}$, a width $\theta_{c}$, and a truncation angle $\theta_{wing}$ outside of which the energy is zero.

\begin{itemize}
    \item Gaussian with Core: a Gaussian with a top-hat core of width $\theta_{c}$.
        \begin{equation}
        E(\theta) = E_{iso} e^{-\frac{\theta^{2}}{2\theta^{2}_{c}}}
        \label{gaussian}
        \end{equation}
    \item Cone: a "hollow" top-hat with constant isotropic-equivalent energy $E_{iso}$ between inner angle $\theta_{c}$ and outer angle $\theta_{wing}$.
        \begin{equation}
          E(\theta) =
          \begin{cases}
                0 & \text{if $\theta < \theta_c$} \\
                E_{iso} & \text{if $\theta_c \leq \theta \leq \theta_{wing}$} \\
          \end{cases}
        \end{equation}
    \item Top Hat: a top-hat with constant isotropic-equivalent energy $E_{iso}$ inside $\theta_{c}$.
        \begin{equation}
          E(\theta) =
          \begin{cases}
                E_{iso} & \text{if $\theta \leq  \theta_c$} \\
                0 & \text{if $\theta > \theta_c$} \\
          \end{cases}
        \end{equation}
\end{itemize}

\subsection{Jet Parameters}

Important parameters that define the structure of a jet and the resulting afterglow light curve include:
\begin{itemize}
    \item $E_{iso}$: Isotropic equivalent kinetic energy (erg)
    \item $n_0$: Circumburst density ($cm^{-3}$)
    \item $\epsilon_{B}$: Fraction of energy in the magnetic fields
    \item $\epsilon_{e}$: Fraction of energy in the electrons 
    \item $p$: Electron energy distribution index
    \item $\theta_{obs}$: Viewing angle (radians)
\end{itemize}

  The latter in particular plays an important role in how we see the light curve and therefore interpret the jet structure. This is where the gravitational wave emission is additionally important as it can provide some constraint on this value.

\section{GW 190814} \label{methods}
The compact object binary merger GW190814 was detected in gravitational waves by Advanced LIGO/Virgo on August 14, 2019. This merger is notable due to the uncertain nature of the binary's lighter-mass counterpart--either the heaviest known neutron star or the lightest known black hole \citep{alexander}. Since two small black holes merging would likely not produce electromagnetic emission, we can help constrain the nature of the merger by searching for electromagnetic emission. \cite{alexander} and \cite{thakur}, among others, have searched near the merger event for a GRB afterglow counterpart of GW190814 in the radio and optical.

\begin{figure*}[htp]
    \centering
    \includegraphics[width=15cm]{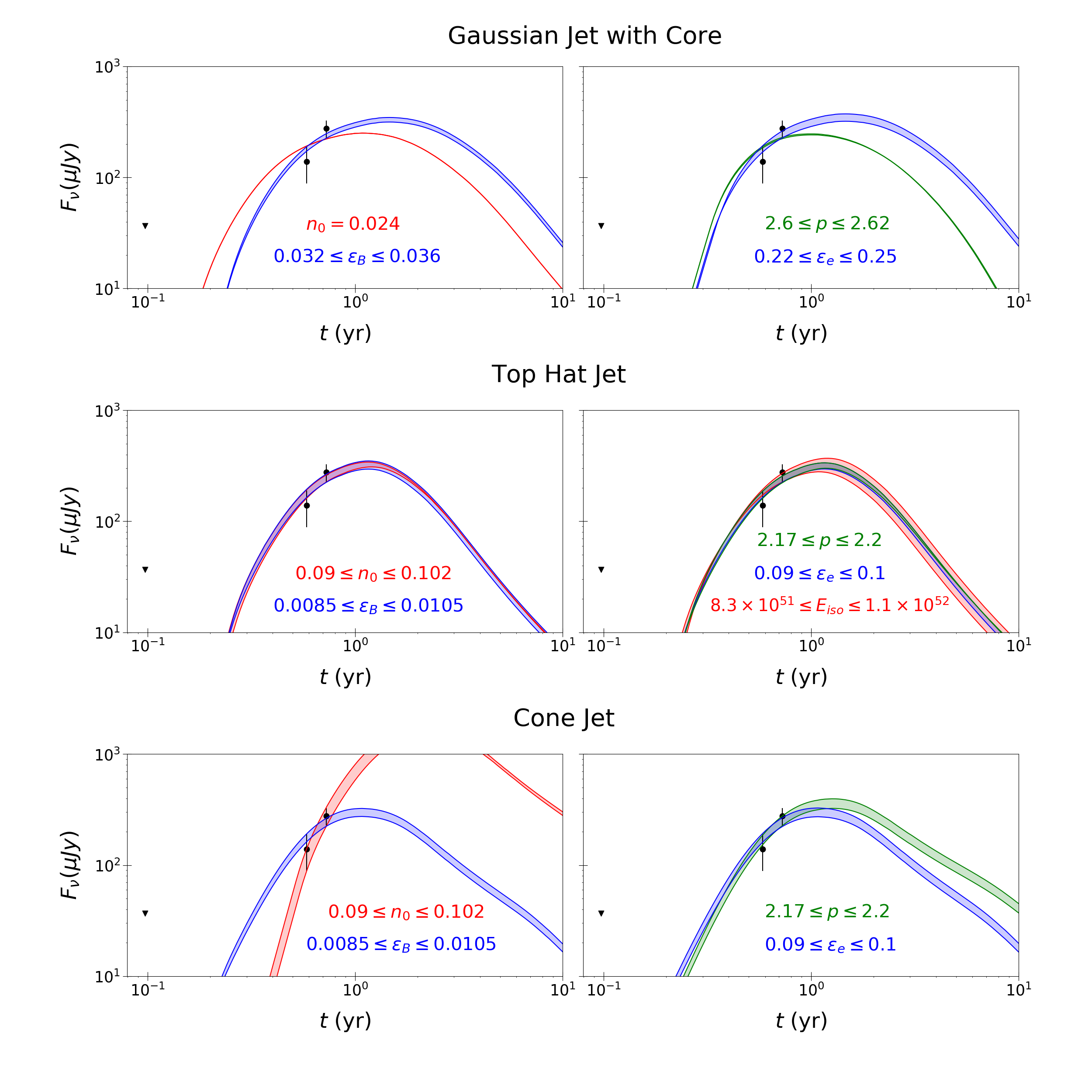}
    \caption{The 5.0 GHz light curve of Candidate 1 (black points, error bars are $1\sigma$). Afterglow light curves of Gaussian jets with cores (top), top hat jets (middle), and cone jets (bottom) with $\theta_{obs} = 46$ degrees and $\theta_{c} = 15$ degrees are fitted to the data. We show the suitable range of $\epsilon_e$,  $\epsilon_B$, $p$, $n_0$, and $E_{iso}$ for each jet structure while all other parameters are held constant (Gaussian: $E_{iso} = 10^{52}$ ergs, $n_{0} = 0.01 cm^{-3}$, $\epsilon_{e} = 0.1$, $\epsilon_{b} = 0.01$, $p = 2.2$; Top Hat: $E_{iso} = 10^{52}$ ergs, $n_{0} = 0.1 cm^{-3}$, $\epsilon_{e} = 0.1$, $\epsilon_{b} = 0.01$, $p = 2.2$; Cone: $E_{iso} = 10^{53}$ ergs, $n_{0} = 0.1 cm^{-3}$, $\epsilon_{e} = 0.1$, $\epsilon_{b} = 0.01$, $p = 2.2$).}
    \label{fig:gaussfit}
\end{figure*}

\cite{alexander} found a promising candidate (Candidate 1) for the radio counterpart of GW190814. Candidate 1 includes a non-detection at 38 days as well as detections at 238 days and 266 days (black triangle and black points, respectively, on Figure \ref{fig:gaussfit}).

Given the viewing angle of $\theta_{obs} = 46^{+17}_{-11}$ degrees from the gravitational wave signal and the distance of $d_{L} = 241^{+41}_{-45}$ Mpc \citep{abbott20}, we produce 6.0 GHz light curves of Gaussian jets with cores, top hat jets, and cone jets that match the flux evolution of Candidate 1 (Figure \ref{fig:gaussfit}).  

 Our results indicate that a number of physically reasonable jet structures fit this data\footnote{\cite{Ben20} show how this degeneracy can be broken to some extent and how jet structure can be better constrained by considering the {\em shape} (rather than simply the flux and time normalizations) of the light curves, when enough data are available.}.  However, a high resolution radio image taken 266 days post merger by Very Large Array (VLA) of this source indicates a double-lobed structure of this source that fairly confidently rules out Candidate 1 as coming from the jet of a GRB associated with GW190814. The emission is partially resolved into two distinct components separated by $\sim 3"$ \citep{alexander2}. At 271 Mpc (the distance of ESO 471-035), that would correspond to a physical separation of $\sim 4$ kpc. This is orders of magnitude too large for a newly formed GRB jet ($<1$ pc). This implies that Candidate 1 is an older and larger source, likely a background double-lobbed radio AGN undergoing a flaring event, although we have not examined the possibility that one lobe is associated with the GRB while the other may be a background source \citep{wagoner}.

\section{Conclusions} \label{Conclusion}
With the end of the third LIGO-Virgo observing run (O3), it becomes particularly urgent to continue the search for electromagnetic counterparts to GW emission, particularly in light of 2 new NS-BH mergers having been discovered \citep{LIGO}. Much of the previous GRB afterglow analysis has relied solely on flux evolution. Yet, this study reveals how closely a background AGN jet could mimic the behavior of a GRB jet and appear to be a GW EM counterpart. Consequently, obtaining high-resolution radio data is crucial in future follow-up efforts, as that may be the only way to distinguish background sources like AGN jets from true GW counterparts.


\section{Acknowledgements}
\label{sec:ack}
We thank Kate Alexander, Wen-fai Fong, Genevieve Schroeder, Kerry Paterson, for very helpful discussions and for sharing their data.  We also are very grateful to Greg Ryan for his help with afterglowpy, and Bob Wagoner for insightful comments and discussions.

\bibliography{main}{}
\bibliographystyle{aasjournal}



\setlength{\arrayrulewidth}{0.5mm}
\setlength{\tabcolsep}{3pt}
\renewcommand{\arraystretch}{1.3}

\end{document}

%% file: header.tex
\usepackage{amsmath}
\usepackage{amssymb}
\usepackage{braket}
\usepackage{amssymb}
\usepackage{color}
\usepackage{xcolor}
\usepackage{graphicx}
\usepackage{latexsym}
\usepackage{listings}
\usepackage{mathrsfs}
\usepackage{letltxmacro}
\usepackage[normalem]{ulem}

\usepackage{subfigure}

\usepackage{verbatim}
\usepackage{tabularx}
\usepackage{ragged2e}
\usepackage{multirow}
\usepackage{amsbsy}

\usepackage{numprint}
\usepackage{makecell}
\usepackage[utf8]{inputenc}

\newcommand{\R}{\mathbb{R}}
\newcommand{\N}{\mathbb{N}} 

\newcommand{\V}{\mathcal{V}}

\newcommand{\F}{\mathcal{F}}

\newcommand{\M}{\mathcal{M}}

\newcommand{\I}{\mathcal{I}}

\newcommand{\B}{\text{B}}

\npdecimalsign{.}
\npthousandsep{,}

\newcolumntype{Y}{>{\RaggedRight\arraybackslash}X}

\graphicspath{{./figures/}}

\sloppypar